\documentclass[amsmath,amssymb,aps,prd,11pt]{revtex4-2}
\usepackage[T1]{fontenc}
\usepackage{xcolor}
\usepackage{graphicx}
\usepackage{mathtools}
\usepackage{ragged2e}

\usepackage{dcolumn}% Align table columns on decimal point
\usepackage{bm}
\usepackage{graphicx,empheq}
\usepackage{verbatim} 
\usepackage[english]{babel}
\usepackage{hyperref}

\hypersetup{
citecolor=red,
colorlinks=true,
filecolor=red,
linkcolor=blue,
linktocpage=true,
urlcolor=blue
} 
\usepackage[titletoc,toc,title]{appendix}
\numberwithin{equation}{section}
\usepackage{cleveref}
\usepackage{epsfig}
\usepackage{amsfonts}
\usepackage{enumitem}
\newcommand\trick[1]{}
\newcommand{\be}{\begin{equation}} 
\newcommand{\ee}{\end{equation}}
\newcommand{\eq}[1]{(\ref{#1})}
\newcommand{\bit}{\begin{itemize}}  \newcommand{\eit}{\end{itemize}}
\newcommand{\ben}{\begin{enumerate}}  \newcommand{\een}{\end{enumerate}}

\newcommand{\rf}[1]{(\ref{#1})}

\def\bd{\begin{document}}
\def\ed{\end{document}}
\def\bea{\begin{eqnarray}}
\def\eea{\end{eqnarray}}
\let\bm=\bibitem

\def\la{\langle}
\def\ra{\rangle}

\def\npb#1#2#3{Nucl. Phys. {\bf{B#1}} #3 (#2)}
\def\plb#1#2#3{Phys. Lett. {\bf{#1B}} #3 (#2)}
\def\prl#1#2#3{Phys. Rev. Lett. {\bf{#1}} #3 (#2)}
\def\prd#1#2#3{Phys. Rev. {D bf{#1}} #3 (#2)}
\def\cmp#1#2#3{Comm. Math. Phys. {\bf{#1}} #3 (#2)}
\def\cqg#1#2#3{Class. Quantum Grav. {\bf{#1}} #3 (#2)}
\def\nppsa#1#2#3{Nucl. Phys. B (Proc. Suppl.) {\bf{#1A}}#3 (#2)}
\def\ap#1#2#3{Ann. of Phys. {\bf{#1}} #3 (#2)}
\def\ijmp#1#2#3{Int. J. Mod. Phys. {\bf{A#1}} #3 (#2)}
\def\rmp#1#2#3{Rev. Mod. Phys. {\bf{#1}} #3 (#2)}
\def\mpla#1#2#3{Mod. Phys. Lett. {\bf A#1} #3 (#2)}
\def\jhep#1#2#3{J. High Energy Phys. {\bf #1} #3 (#2)}
\def\atmp#1#2#3{Adv. Theor. Math. Phys. {\bf #1} #3 (#2)}

\def\sst{\scriptscriptstyle}
\def\thetabar{\bar\theta}
\def\Tr{{\rm Tr}}
\def\one{\mbox{1 \kern-.59em {\rm l}}}

\def\a{\alpha}      \def\da{{\dot\alpha}}  \def\dA{{\dot A}}
\def\b{\beta}       \def\db{{\dot\beta}}
\def\g{\gamma}  \def\G{\Gamma}  \def\dc{{\dot\gamma}}
\def\d{\delta}  \def\D{\Delta}  \def\ddt{\dot\delta}
\def\e{\epsilon}
\def\ve{\varepsilon}
\def\uve{\upvarepsilon}
\def\f{\phi}    \def\F{\Phi}    \def\vvf{\f}
\def\vphi{\varphi}
\def\h{\eta}
\def\k{\kappa}
\def\l{\lambda} \def\L{\Lambda}
\def\m{\mu} \def\n{\nu}
\def\o{\omega}
\def\p{\pi} \def\P{\Pi}
\def\r{\rho}
\def\s{\sigma}  \def\S{\Sigma}
\def\t{\tau}
\def\th{\theta} \def\Th{\Theta} \def\vth{\vartheta}
\def\X{\Xeta}
\def\z{\zeta}

\def\na{\nabla}
\def\cA{{\cal A}} \def\cB{{\cal B}} \def\cC{{\cal C}}
\def\cD{{\cal D}} \def\cE{{\cal E}} \def\cF{{\cal F}}
\def\cG{{\cal G}} \def\cH{{\cal H}} \def\cI{{\cal I}}
\def\cJ{{\mathrm J}} \def\cK{{\cal K}} \def\cL{{\cal L}}
\def\cM{{\cal M}} \def\cN{{\cal N}} \def\cO{{\cal O}}
\def\cP{{\cal P}} \def\cQ{{\cal Q}} \def\cR{{\cal R}}
\def\cS{{\cal S}} \def\cT{{\cal T}} \def\cU{{\cal U}}
\def\cV{{\cal V}} \def\cW{{\cal W}} \def\cX{{\cal X}}
\def\cY{{\cal Y}} \def\cZ{{\cal Z}}
\def\ct{{\cal t}}

\def\ua{\underline{\alpha}}
\def\uc{\underline{\phantom{\alpha}}\!\!\!\gamma}
\def\um{\underline{\mu}}
\def\ud{\underline\delta}
\def\ue{\underline\epsilon}
\def\una{\underline a}\def\unA{\underline A}
\def\unb{\underline b}\def\unB{\underline B}
\def\unc{\underline c}\def\unC{\underline C}
\def\und{\underline d}\def\unD{\underline D}
\def\une{\underline e}\def\unE{\underline E}
\def\unf{\underline{\phantom{e}}\!\!\!\! f}\def\unF{\underline F}
\def\unm{\underline m}\def\unM{{\underline M}}
\def\unn{\underline n}\def\unN{{\underline N}}
\def\unp{\underline{\phantom{a}}\!\!\! p}\def\unP{\underline P}
\def\unq{\underline{\phantom{a}}\!\!\! q}
\def\unQ{\underline{\phantom{A}}\!\!\!\! Q}
\def\unH{\underline{H}}

\def\As {{A \hspace{-6.4pt} \slash}\;}
\def\bs {{b \hspace{-6.4pt} \slash}\;}
\def\Ds {{D \hspace{-6.4pt} \slash}\;}
\def\Gts {{\Gt \hspace{-6.4pt} \slash}\;}
\def\ds {{\del \hspace{-6.4pt} \slash}\;}
\def\ss {{\s \hspace{-6.4pt} \slash}\;}
\def\ks {{ k \hspace{-6.4pt} \slash}\;}
\def\ps {{p \hspace{-6.4pt} \slash}\;}
\def\xs {{x \hspace{-6.4pt} \slash}\;}
\def\pas {{{p_1} \hspace{-6.4pt} \slash}\;}
\def\pbs {{{p_2} \hspace{-6.4pt} \slash}\;}
\def\cFs {{{\cal F} \hspace{-6.4pt} \slash}\;}
\def\Dss {{D \hspace{-7.5pt} \slash}\;}
\def\dss {{\del \hspace{-7.0pt} \slash}\;}

\def\Ah{{\hat{A}}}
\def\Dh{{\hat{D}}}
\def\Gh{{\hat{G}}}
\def\Fh{{\hat{F}}}
\def\Ih{{\hat{I}}}
\def\Jh{{\hat{J}}}
\def\Kh{{\hat{K}}}
\def\Lh{{\hat{L}}}
\def\Ph{{\hat{P}}}
\def\Rh{{\hat{R}}}
\def\Vh{{\hat{V}}}
\def\Xh{{\hat{X}}}

\def\ah{{\hat{\a}}}
\def\bh{{\hat{\b}}}
\def\gh{{\hat{\g}}}
\def\dh{{\hat{\d}}}
\def\rh{{\hat{\r}}}
\def\hh{\hat{h}}
\def\uh{\hat{u}}
\def\xh{\hat{x}}
\def\yh{\hat{y}}
\def\ph{\hat{p}}
\def\xih{\hat{\xi}}
\def\chih{\hat{\chi}}
\def\Psih{\hat{\Psi}}
\def\phih{\hat{\phi}}

\def\psit{\tilde{\psi}}
\def\Psit{\tilde{\Psi}}
\def\Psibt{\tilde{\bar{Psi}}}

\def\lambdat{\tilde {\lambda}}
\def\st{\tilde{\sigma}}

\def\delt{\tilde{\delta}}
\def\Phit{\tilde{\Phi}}
\def\Phitb{\overline{\tilde{Phi}}}
\def\tht{\tilde{\th}}
\def\lt{\tilde{\l}}
\def\chit{\tilde{\chi}}
\def\phit{\tilde{\phi}}

\def\At{\tilde{A}}
\def\Bt{\tilde{B}}
\def\Ct{\tilde{C}}
\def\Dt{\tilde{D}}
\def\Et{\tilde{E}}
\def\Ft{\tilde{F}}
\def\Gt{\tilde{G}}
\def\Ht{\tilde{H}}
\def\It{\tilde{I}}
\def\Jt{\tilde{J}}
\def\Qt{\tilde{Q}}
\def\Rt{\tilde{R}}
\def\Mt{\tilde{M }}
\def\Nt{\tilde{N}}
\def\St{\tilde{S}}
\def\Vt{\tilde{V}}
\def\Xt{\tilde{X}}
\def\at{\tilde{a}}
\def\dt{\tilde{d}}
\def\htt{\tilde{h}}
\def\ft{\tilde{f}}
\def\gt{\tilde{g}}
\def\pt{\tilde{p}}
\def\qt{\tilde{q}}
\def\vt{\tilde{v}}
\def\nt{\tilde{n}}
\def\ut{\tilde{u}}
\def\wt{\tilde{w}}
\def\zt{\tilde{z}}
\def\xt{\tilde{x}}
\def\yt{\tilde{y}}
\def\Psit{\tilde{\Psi}}
\def\vphit{\tilde{\varphi}}
\def\tD{\tilde{\D}}

\def\eb{\bar{\epsilon}}
\def\delb{\bar{\partial}}
\def\thb{\bar{\theta}}
\def\mub{\bar{\mu}}
\def\lamb{\bar{\l}}
\def\psib{\bar{\psi}}
\def\sb{\bar{\sigma}}
\def\xib{\bar{\xi}}
\def\chib{\bar{\chi}}

\def\Psib{\bar{\Psi}}
\def\Phib{\bar{\Phi}}
\def\Lamb{\bar{\Lambda}}
\def\Sb{{\overline \Sigma}}
\def\cb{\bar{c}}
\def\hb{\bar{h}}
\def\qb{\bar{q}}
\def\wb{\bar{w}}
\def\ub{\bar{u}}
\def\zb{{\bar{z}}}
\def\Hb{\bar{H}}
\def\Qb{{\bar Q}}
\def\Omegab{\overline{\Omega}}
\def\ob{\overline{\omega}}

\def\Ab{{\overline A}} \def\Bb{{\overline B}} \def\Cb{{\overline C}}
\def\Db{{\overline D}} \def\Eb{{\overline E}} \def\Fb{{\overline F}}
\def\Gb{{\overline G}}
\def\Ib{{\overline I}}
\def\Jb{{\overline J}} \def\Kb{{\overline K}} \def\Lb{{\overline L}}
\def\Mb{{\overline M}} \def\Nb{{\overline N}} \def\Ob{{\overline O}}
\def\Pb{{\overline P}}  \def\Rb{{\overline R}}
 \def\Tb{{\overline T}} \def\Ub{{\overline U}}
\def\Vb{{\overline V}} \def\Wb{{\overline W}} \def\Xb{{\overline X}}
\def\Yb{{\overline Y}} \def\Zb{{\overline Z}}

\def\fb{{\overline f}}
\def\gb{{\overline g}}
\def\nb{{\overline n}}
\def\mb{{\overline m}}
\def\lb{{\overline l}}
\def\yb{{\overline y}}

\def\ldel{{\overleftarrow{\del}}}
\def\rdel{{\overrightarrow{\del}}}
\def\ldeldel{{\overleftarrow{\del^2}}}
\def\rdeldel{{\overrightarrow{\del^2}}}
\def\ldelb{{\overleftarrow{\bar{\del}}}}
\def\rdelb{{\overrightarrow{\bar{\del}}}}
\def\ba{{\bf a}}
\def\bk{{\bf k}}
\def\bl{{\bf l}}
\def\bp{{\bf p}}
\def\bq{{\bf q}}
\def\br{{\bf r}}
\def\bt{{\bf t}}
\def\bu{{\bf u}}
\def\bv{{\bf v}}
\def\bx{{\bf x}}
\def\by{{\bf y}}
\def\bA{{\bf A}}
\def\bR{{\bf R}}
\def\bV{{\bf V}}

\def\bz{{\boldsymbol{\zeta}}}

\def\bone{{\bf 1}}

\def\va{{\vec a}}
\def\vk{{\vec k}}
\def\vp{{\vec p}}
\def\vq{{\vec q}}
\def\vx{{\vec x}}
\def\vy{{\vec y}}
\def\vu{{\vec u}}
\def\vv{{\vec v}}
\def \vH{{\vec H}}
\def \vg{{\vec g}}

\def\vs{{\vec \sigma}}
\def\vtau{{\vec \tau}}

\newcommand{\ov}[1]{\overrightarrow{#1}}

\def\frA{\mathfrak{A}}
\def\frB{\mathfrak{B}}
\def\frC{\mathfrak{C}}
\def\frD{\mathfrak{D}}
\def\frE{\mathfrak{E}}
\def\frF{\mathfrak{F}}
\def\frG{\mathfrak{G}}
\def\frH{\mathfrak{H}}
\def\frM{\mathfrak{M}}
\def\frN{\mathfrak{N}}
\def\frR{\mathfrak{R}}
\def\frW{\mathfrak{W}}

\def\fra{\mathfrak{a}}
\def\frb{\mathfrak{b}}
\def\frf{\mathfrak{f}}
\def\frg{\mathfrak{g}}
\def\frh{\mathfrak{h}}
\def\frl{\mathfrak{l}}
\def\frs{\mathfrak{s}}
\def\fri{\mathfrak{i}}
\def\frj{\mathfrak{j}}

\def\ma{\mathfrak{a}}
\def\mg{\mathfrak{g}}
\def\mh{\mathfrak{h}}
\def\mR{\mathfrak{R}}
\def\mN{\mathfrak{N}}

\newcommand{\nn}{{\nonumber}}

\def\d{\delta}\def\D{\Delta}\def\ddt{\dot\delta}

\def\pa{\partial} \def\del{\partial}
\def\xx{\times}
\def\uno{\mbox{1 \kern-.59em {\rm l}}}

\def\trp{^{\top}}
\def\inv{^{-1}}
\def\dag{\dagger}
\def\pr{^{\prime}}

\def\rar{\rightarrow}
\def\lar{\leftarrow}
\def\lrar{\leftrightarrow}

\newcommand{\0}{\,\!}      %this is just NOTHING!
\def\one{1\!\!1\,\,}
\def\im{\imath}
\def\jm{\jmath}

\newcommand{\tr}{\mbox{tr}}
\newcommand{\slsh}[1]{/ \!\!\!\! #1}

\newcommand{\1}{\mbox{1}\hspace{-0.25em}\mbox{l}}

\def\vac{|0\rangle}
\def\lvac{\langle 0|}

\def\hlf{\frac{1}{2}}
\def\ove#1{\frac{1}{#1}}
\newcommand{\hot}[1]{\frac{#1}{2}}

\def\Box{\square}
\def\CC {\mathbb{C}}
\def\FF {\mathbb{F}}
\def\RR{\mathbb{R}}
\def\NN{\mathbb{N}}
\def\ZZ{\mathbb{Z}}
\def\bb#1{{\bf #1}}
\def\bcomment#1{}
\def\bfhat#1{{\bf \hat{#1}}}
\def\VEV#1{\left\langle #1\right\rangle}

\newcommand{\ex}[1]{{\rm e}^{#1}} \def\ii{{\rm i}}

\newcommand{\lrbrk}[1]{\left(#1\right)}
\newcommand{\lrsbrk}[1]{\left[#1\right]}
\newcommand{\sfrac}[2]{{\textstyle\frac{#1}{#2}}}

\def\stw{{\sqrt{2}}}

\def\rf {{\rm f}}
\def\ri {{\rm i}}
\def\rj {{\rm j}}
\def\rn {{\rm n}}
\def\rk {{\rm k}}
\def\rl {{\rm l}}
\def\rr {{\rm r}}
\def\rs {{\scriptscriptstyle \rm S}}
\def\rt {{\scriptscriptstyle \rm T}}

\def\rQ {{\scriptscriptstyle \rm \cQ}}
\def\rR {{\scriptscriptstyle \rm \cR}}

\def\cQb{{\cal \Qb}}
\def\cRb{{\cal \Rb}}
\def\cWb{{\cal \Wb}}

\def\fd {{\rm N}}
\def\afd {{\overline{\rm N}}}

\def \II {I\hspace{-.1em}I\hspace{.1em}}
\def \IIA {\mbox{\II A\hspace{.2em}}}
\def \IIB {\mbox{\II B\hspace{.2em}}}
\def \gs {g^s}
\def \ls {\lambda^s}

\def \I {{\cal I}}
\def \qs {q\hspace{-.53em}/\hspace{.15em}}
\def \ks {k\hspace{-.53em}/\hspace{.15em}}
\def \YM {{\mbox{\tiny YM}}}
\def \gym {g_{\YM}}

\def \Lc {\L_c}
\def\IR{\relax{\rm I\kern-.18em R}}
\def \id {{\bf 1}}

\def\cci{\ell}
\def\ccj{\ell'}

\def\bbq{\pmb{q}}
\def\bom{\pmb{\o}}
\def\bJ{\pmb{J}}
\def\bM{\pmb{M}}
\def\bB{\pmb{B}}
\def\bn{\pmb{n}}
\def\bE{\pmb{E}}

\newcommand{\rrr}[1]{\vskip 0.2cm \noindent{\bf #1} ---}

\long\def\symbolfootnote[#1]#2{\begingroup%
\def\thefootnote{\fnsymbol{footnote}}\footnote[#1]{#2}\endgroup}
\long\def\RemarkBox#1{\begin{flushleft}\fbox{\begin{minipage}
{17.5cm}{\bf Remark:} ~#1\end{minipage}}\end{flushleft}}
\newcommand{\aei}{\it Max Planck Institute for Gravitational Physics
(Albert Einstein Institute)\\ Am M\"uhlenberg 1, 14476 Golm,
Germany}

\newcommand{\nthu}{{\it Department of Physics, National Tsing-Hua
  University,
  Hsinchu 30013, Taiwan}}

\newcommand{\ctc}{{\it
Center of Theory and Computation, 
National Tsing-Hua University, Hsinchu 30013, Taiwan}}

\newcommand{\ncts}{{\it
National Center for Theoretical Sciences, Taipei 10617, Taiwan}}

\begin{document}
\title{Lifshitz Quantum Mechanics and Anisotropic Josephson Junction}
\author{Chong-Sun Chu${}^{1,2,3}$ }
\email[Correspondence email address: ]{cschu@phys.nthu.edu.tw}
\author{Alfian Gunawan${}^1$}
\affiliation{${}^1$ Department of Physics, National Tsing-Hua
  University,  Hsinchu 30013, Taiwan}
\affiliation{${}^2$ Center of Theory and Computation, 
National Tsing-Hua University, Hsinchu 30013, Taiwan}
\affiliation{${}^3$
  National Center for Theoretical Sciences, Taipei 10617, Taiwan}

\begin{abstract}  
  We consider
  %c4  Lifshitz
  quantum mechanics in spacetime with anisotropy in time. Such Lifshitz
  quantum mechanics is characterized by  a kinetic term with
  fractional derivatives.
  We show that, contrary to
  %c5 popular belief,
  a common claim in the literature,
  local conservation of probability is respected when the probability
  current is properly identified.
  As an  application we consider a
  Josephson Junction with
  %c5
  an insulating layer exhibiting Lifshitz anisotropy.
  We show that
  anisotropy modifies the tunneling rate and can
  %c5 greatly
  significantly enhance the performance of the  Josephson Junction.

\end{abstract}

\maketitle

%\tableofcontents

%\setcounter{footnote}{0}

\section{Introduction}

Many interesting systems in physics are anisotropic.
For example,
liquid crystals, magnetic materials and critical systems
in condensed matter physics are ubiquitously anisotropic. 
The cosmic microwave background
also display traces of anisotropy that are thought to have originated from
quantum fluctuations occurred during inflation. 
Often the fundamental physical laws are isotropic but the
%c5 presence of a
background structure
%c5 leads to
induces anisotropy. 
An interesting instance of anisotropy is
when physics is invariant under the  Lifshitz scaling transformation
\begin{equation}
t\rightarrow \lambda^z t, \quad  x^i \rightarrow \lambda x^i ,\quad
   \l>  0.
	\label{lss}
\end{equation}
Here spatial isotropy is maintained while time is anisotropic for
%c5 an anisotropy
a dynamical 
exponent $z \neq 1$.
Such Lifshitz scaling symmetry occurs
naturally as the symmetry at the critical point
of renormalization group (RG) flow for anisotropic systems. 
Lifshitz scale invariant system can be recognized easily by the dispersion
relation of their massless modes:
\be \label{dispersion}
\o \sim  k^{z} ,
\ee
%c5 where
from which one can extract not only the dynamical exponent $z$
but also a mass scale associated with the anisotropy. As the only possible
form that is compatible with the scaling symmetry \eq{lss},
\eq{dispersion} is  independent of the
underlying origin of the anisotropy.
%c4
The symmetry for $z=2$
was first proposed by Lifshitz \cite{lif} to describe the
tricritical point of liquid helium \cite{Hornreich:1975zz}.
The dispersion relation \eq{dispersion}
%c5 is also featured 
also appears in certain  modified gravity scenarios involving
ghost condensate 
\cite{Arkani-Hamed:2003pdi}. In condensed matter physics,
a well known  example
is the  Rokhsar-Kivelson (RK) Quantum dimer
model \cite{PhysRevLett.61.2376,Henley_2004}
where a
continuum $z=2$ Lifshitz scalar field theory emerges at the critical point.

%c5
Explicit field theoretic realization of the Lifshitz symmetry \eq{lss}
for the special integer value of
$z=2$ has been known as the quantum Lifshitz model (QLM)
\cite {Ardonne_2004} . Various
bipartite entanglement measures were subsequently analyzed within the
QLM
\cite{
Fradkin:2006mb,Fradkin:2009dus, PhysRevB.80.184421,
  Hsu:2008af,Hsu:2010ag,Oshikawa:2010kv,
  PhysRevLett.107.020402,Zhou:2016ykv,MohammadiMozaffar:2017nri,
  Berthiere:2019lks, Angel-Ramelli:2019nji,
  Berthiere:2019lks, Angel-Ramelli:2020wfo, Berthiere:2023bwn}.
It is known that
Lifshitz invariant scalar field theory admits ground state called the
Rokhsar-Kivelson (RK) vacuum \cite{PhysRevLett.61.2376, Henley_2004},
with the properties that it
is local and is given by a superposition of
quantum states determined by some classical probability distribution.

%c5
The point $z=2$  is however
somewhat special since the Lifshitz scaling transformation \eq{lss}
combines with the Galilean boost to form
the Schrodinger group only for this value of $z$.
The study of field theory beyond $z=2$ was pioneered in \cite{Keranen:2016ija} where 
QLM was  generalized to $(D+1)$-dimensions with $z=D$. It was  also
realized very interestingly that the connection
to conformal field theory (CFT) found in QLM is retained for general $z=D>2$.
This was further generalized in \cite{Basak:2023otu} where
field theory  for general non-integer $z$ was constructed
with the help of fractional derivatives. It was found that
the ground state of the  Lifshitz invariant scalar field continue to take 
the form of RK for general $z$. Furthermore, it was found that
even though the usual CFT twisted operator is not available in the Lifshitz invariant
field theory, due to the special property of the RK vacuum,
it is possible to employ the replica trick in a path integral formalism that they developed
to study
the entropic properties of these theories \cite{Basak:2023otu}. It was found that
the results are connected to  CFT for more general non-integer value of $z$. Further results
in holography were also obtained in \cite{Basak:2023otu}, and a holographic
proposal for boundary Lifshitz field theory in \cite{Chu:2024nwf}.

Clearly, the physics of anisotropy is a subject of its own interest that
goes beyond  quantum field theory. With general applicability in mind,
we explore in this paper
the physical effect of anisotropy in the more
basic quantum mechanics. In particular
we consider  time anisotropy of the background space time specified by \eq{lss}.
With the dispersion relation fixed to be of the form \eq{dispersion},
quantum mechanics of free particle 
in the anisotropic spacetime reads
\be \label{H0}
i \hbar \frac{\del}{\del t} \psi = H_0 \psi, \quad
H_0 = -\frac{\hbar^z}{M^{z-2}}\frac{\D_z}{2m}
\ee
where $H_0$ is the free Hamiltonian, $m$ is the mass of the particle
and $M$ is a mass scale associated with the anisotropy.
Here, defined by its action 
\be \label{fracD}
\del_j^{z} e^{i \bk \cdot \bx} := (i k_j)^z e^{i \bk \cdot \bx}
\ee
on the basis of plane wave functions, $\del_j^{z}$
is an fractional derivative of order $z$, and
$\D_z : = \del_j ^z \del_j^z$  is  the fractional Laplacian
\begin{eqnarray} \label{fracD0}
    \triangle_z \, e^{i \bk \cdot \bx} = (- \bk^2)^{z/2} e^{i \bk\cdot \bx}.
\end{eqnarray}
The free Hamiltonian \eq{H0} is Lifshitz scale invariant. 
One may also add a potential and consider the
{\it Lifshitz quantum mechanics} defined by
\be \label{H}
i \hbar \frac{\del}{\del t} \psi = H \psi, \quad
H = -\frac{\hbar^z}{M^{z-2}}\frac{\D_z}{2m} + V.
\ee
%c5
Note that other definitions of fractional derivatives are possible.
For example, the Riesz derivative (see e.g. \cite{Laskin:2010ry})
is defined also in terms of the Fourier
transformation. However, non-analytic function $|k_j|$
is involved and it does not allow one to use complex contour integral
theorems.  For the analysis of   \cite{Basak:2023otu}, it was crucial to  adopt
the definition \eq{fracD} since results such as
the solution $\phi = x^{z-n}$  to the equation of motion $\del_x^z \phi =0$
was obtained only with the help of  contour deformation techniques.
In this paper, we are interested in quantum mechanics that can
be obtained as a low energy single particle  limit of the Lifshitz QFT,
therefore we consider the form \eq{H} of Lifshitz quantum
mechanics instead of fractional quantum mechanics defined
by other types of fractional derivatives (see e.g. \cite{Laskin:2010ry}).
Note also that in general the scaling symmetry is broken  except for
some special form of the potential,
%c4
e.g. $V \sim 1/|x|^z$, 
in which case we obtain the
Lifshitz invariant quantum mechanics.
%c3
In analogous to conformal quantum mechanics \cite{deAlfaro:1976vlx},
it may be interesting to study further this kind of
Lifshitz invariant quantum mechanics.

Here is a summary of the paper.
In section 2, we consider the Lifshitz quantum mechanics in general.
There is a  belief in the literature that local conservation
of probability is violated by a source term in
%c5 Lifshitz
fractional 
quantum mechanics.
%c4
We show that this is not true for the Lifshitz quantum mechanics.
%c5 It turns out it is
In fact here  the  probability current is modified by anisotropy,
and once  the correct probability current is
employed, the local conservation  of probability is respected.
We argue that similarly
a careful identification of the probability current is needed for
the other types of fractional quantum mechanics and probability is
conserved in fractional quantum mechanics in general. 
In section 3, we consider an anisotropic
Josephson Junction that is made up of an anisotropic insulating layer
sandwiched between two normal superconducting materials. We
show that the critical current can be significantly
enhanced
by anisotropy. We argue that anisotropic Josephson Junction has great
potential in quantum computing.

%c4
\section{Lifshitz    Quantum Mechanics}

Consider Lifshitz quantum mechanics \eq{H}.
This kind of quantum mechanics is also refereed to as fractional
quantum mechanics, see for example \cite{Laskin:2010ry}
for some discussion.
It is straightforward to show that the Lifshitz
quantum mechanics \eq{H} is unitary and
admits the local conservation law
\be \label{cont}
\frac{\del \rho}{\del t} + \nabla \cdot \bJ =0,
\ee
where the probability density takes the usual form
$\rho =\psi^* \psi$, while the
probability current density is modified by anisotropy.
    Since the fractional derivative is defined using
    the basis of exponential functions, $\bJ$ can be most
    easily obtained in the momentum space and we have
  \be \label{Jp}
%c3
  \bJ = - \frac{1}{2m M^{z-2}}
\int \frac{d^D{p}\, d^D{q}}{(2\pi \hbar)^D} \,
e^{i (\bp-\bq)\cdot\bx/\hbar}
      \varphi^*(\bq,t)\varphi(\bp,t)
      \frac{(-\bq^2)^{z/2}-(-\bp^2)^{z/2}}{\bq^2- \bp^2}
      (\bq+\bp),
    \ee
    where
    $ \varphi(\bp,t)$ is the momentum space wave function
    defined by the Fourier transform
\be
    \psi(\bx,t) =
    \int \frac{d^D p}{(2\pi \hbar)^{D/2}} \, e^{i \bp\cdot \bx/
      \hbar}
    \varphi(\bp,t).
\ee
This result can be obtained from the Noether theorem
    associated with    the phase symmetry
    $\psi \to \psi e^{i\th}$
    of quantum mechanics
    or by the direct manipulation of the wave equation \eq{H}.
    %c5
    In fact, it is
    \be
    \psi^* \D_z \psi - \psi \D_z \psi^* = \int_p \int_q
    e^{i (\bq -\bp) \cdot \bx} \varphi^*(\bp) \varphi(\bq)
    ((-\bq^2)^{z/2}-(-\bp^2)^{z/2} ),
    \ee
    where we have denoted $\int_p :=  \int \frac{d^D p}{(2\pi \hbar)^{D/2}}$.
    This can be identify this with a divergence term $\nabla \cdot {\bf K}$
  of the form
    \be
       {\bf K} = -i \int_p \int_q {\bf b} (\bp,\bq) e^{i (\bq -\bp) \cdot \bx}
\varphi^*(\bp) \varphi(\bq) ((-\bq^2)^{z/2}-(-\bp^2)^{z/2} )
\ee
if ${\bf b}$ satisfies
\be \label{bpq}
{\bf b}\cdot (\bq -\bp) =1.
\ee
Here ${\bf K}$ is the current density up to a normalization factor.
The solution to \eq{bpq} is not unique in 3  spatial dimensions. 
For example, \eq{bpq} can be solved by (with a suitable
 regulator implemented, e.g. $i \ve$ in the denominator),
\be \label{egbpq}
   {\bf b} = \frac{\bq +\bp}{\bq^2 - \bp^2}, \quad \mbox{or}\quad
    {\bf b} = \frac{\bq -\bp}{|\bq -\bp|^2},
    \ee
It is easy to understand the physical
meaning of the non-uniqueness. Consider any two solutions
${\bf b_1}, {\bf b_2}$ of \eq{bpq},
then by definition, ${\bf b_2}- {\bf b_1}$ is perpendicular to $\bq -\bp$.
    This means
    \be
    {\bf b_2}- {\bf b_1} = (\bq -\bp) \times {\bf s}
    \ee
    for some vector ${\bf s(\bp, \bq)}$.
    This then implies that the corresponding difference
    ${\bf K_2}- {\bf K_1}$ is given by a curl:
    \be
       {\bf K_2}- {\bf K_1} = - \nabla \times \int_p \int_q
       {\bf s} \; e^{i (\bq -\bp) \cdot \bx}
\varphi^*(\bp) \varphi(\bq) ((-\bq^2)^{z/2}-(-\bp^2)^{z/2} ).
    \ee
    Therefore the different solutions of \eq{bpq} simply give rises to
    current densities that are different from each other by a curl term. This
    does not affect the expression of current conservation.
    For example, for
    the first solution in \eq{egbpq}, we obtain in coordinate space, 
\be \label{Jx}
     \mathbf{J}
    =  \l_M^{z-2} \frac{\hbar}{2mi} 
    \frac{\nabla_x-\nabla_{x'}}{\nabla_x^2-\nabla_{x'}^2}
    \left[(\nabla_x^2)^{z/2}-(\nabla_{x'}^2)^{z/2}\right]
    \psi^*(\mathbf{x}',t)\psi(\mathbf{x},t)
    \Big|_{\mathbf{x}'=\mathbf{x}},
    \ee
where 
%c3
$\l_M : = \hbar/M$ is a length scale associated with the anisotropy.
%c5
We remark that this kind of
ambiguity is there even for the standard quantum
mechanics although it is not usually mentioned.
%c4 
Like the Poynting vector, additional input from the physical setting such as 
symmetry, boundary condition
or  a good understanding of the underlying dynamics, is
usually needed to fix the choice of $\bJ$ properly.
%c5 This is the case for the example that we will consider below.
%c5
We remark also that the local conservation of probability
in fractional quantum mechanics has been discussed
in \cite{Tayurskii2012, Wei2016, Laskin2016}
where a source term has been found, and
which has led to the
%c5 confusion
conclusion of a non-conservation
of probability. If this is indeed the case, it posts serious
doubt in the physical consistency of the fractional quantum mechanics.
However as we showed above,
the probability current is indeed modified in the Lifshitz
%c5 fractional
quantum mechanics and  local conservation of probability is observed. 
We believe that the probability current
considered in the fractional quantum mechanics
\cite{Tayurskii2012, Wei2016, Laskin2016} can also be improved similarly 
so as to maintain probability conservation in quantum mechanics.
%c5
This point is important in order to have a consistent physical interpretation of the
mathematical results computed from the fractional differential equations.
%c5
In the next section, we consider Lifshitz quantum mechanics for a  Josephson Junction system and predict an
observable effect on the macroscopic supercurrent from 
the conserved probability current in the  insulating layer of the Josephson Junction.

\section{Anisotropic Josephson Junction}

As an example of how anisotropy may lead to interesting quantum physics,
we consider tunneling through an anisotropic region. A prominent
example where the tunneling effect has a macroscopic manifestation is the
Josephson effect where cooper pairs of electrons can flow between two
superconductors separated by a thin insulating barrier.

Consider a SIS type
Josephson Junction that is made up of two superconductors separated by a thin
insulator. The two superconductors are assumed to be made of
ordinary isotropic materials while the insulator is made of
an anisotropic material.
The latter can be achieved by material engineering such that
the dispersion relation \eq{dispersion} arises and the insulator
behaves effectively as an anisotropic material.   
In quantum mechanics, the superconductors have Landau-Ginzburg order parameters
\be \label{LGOP}
\psi_i = \sqrt{n_i} e^{\ri \th_i}, \quad i =1,2
\ee
that can be interpreted as the wave functions of the Cooper pairs in the two
superconducting regions. See figure 1.
\begin{figure}
			\centering
                        \includegraphics[scale=0.25]{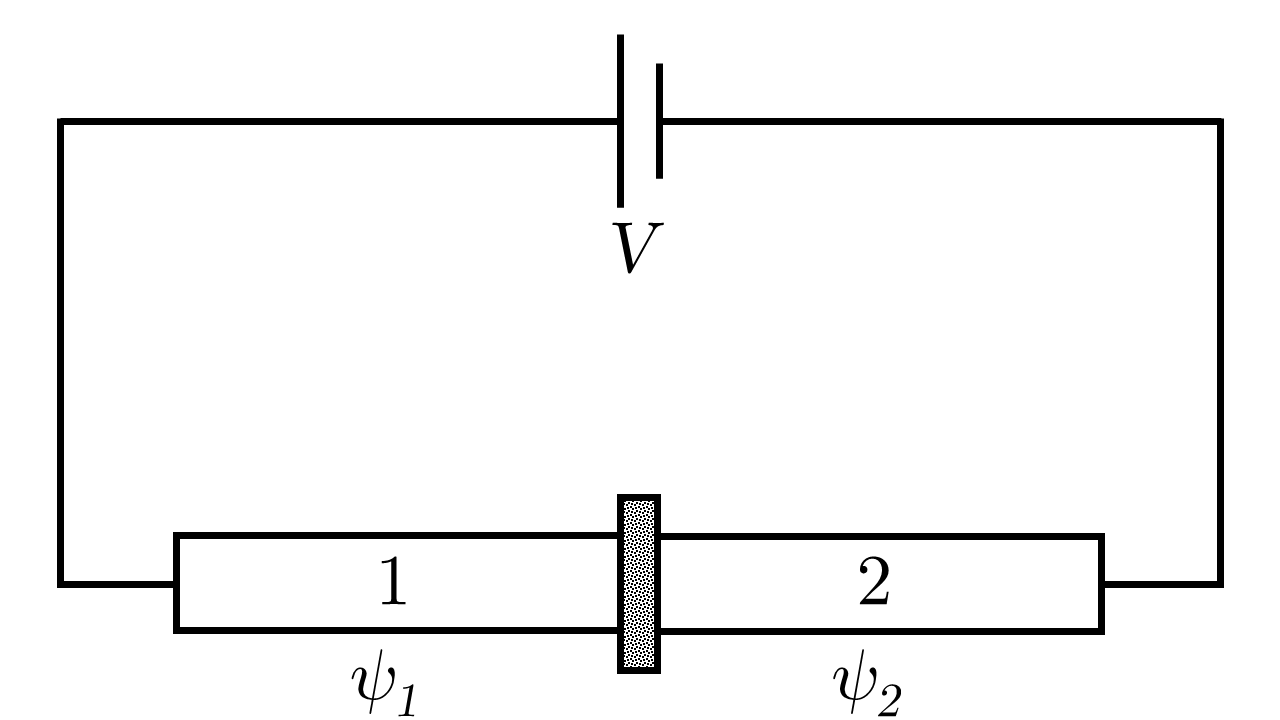}
			\caption{Two superconductors separated by a thin
                          insulator}
			\label{fig1}
\end{figure}
Suppose the Cooper pairs have typical energy $E$
in the regions 1, 2.
%c5
For a junction with planar symmetry and that the insulator may be approximated
by a potential
barrier $V_0 >E$ (see figure 2), the wave equation  inside the insulator
reads
\be
\l_M^{z-2} \frac{\hbar^2
  %c5 \D_z
  (\del_x^{z})^2\psi}{2m}
= (V_0 -E) \psi, \quad -a <x<a, 
\ee
where $x$ is the direction perpendicular to the plane of the layers. Using the definition
\eq{fracD} of the fractional derivative, we obtain the general solution
\be
\psi = C_1 \cosh \frac{x}{\zeta} + C_2 \sinh \frac{x}{\zeta} ,
\ee
where
$C_1, C_2$ are  arbitrary constants  and
\be \label{zeta}
\zeta := \zeta_0 \left(\frac{\zeta_0}{\l_M}\right)^{2/z-1},
\quad \mbox{with}\;\;  \zeta_0 := \frac{\hbar}{\sqrt{2m(V_0-E)}}.
\ee
%c3
Here $\zeta$ is the tunneling penetration length
and $\zeta_0$ is the  tunneling penetration length in the
%c5 undeformed $z=2$ (isotropic)
standard Schrödinger $(z=2)$ case. Note that the penetration length is modified
by anisotropy through the presence of the anisotropy
length scale $\l_M$ and anisotropy index $z$. Next let us determine
the tunneling current.
\begin{figure}
			\centering
                        \includegraphics[scale=0.25]{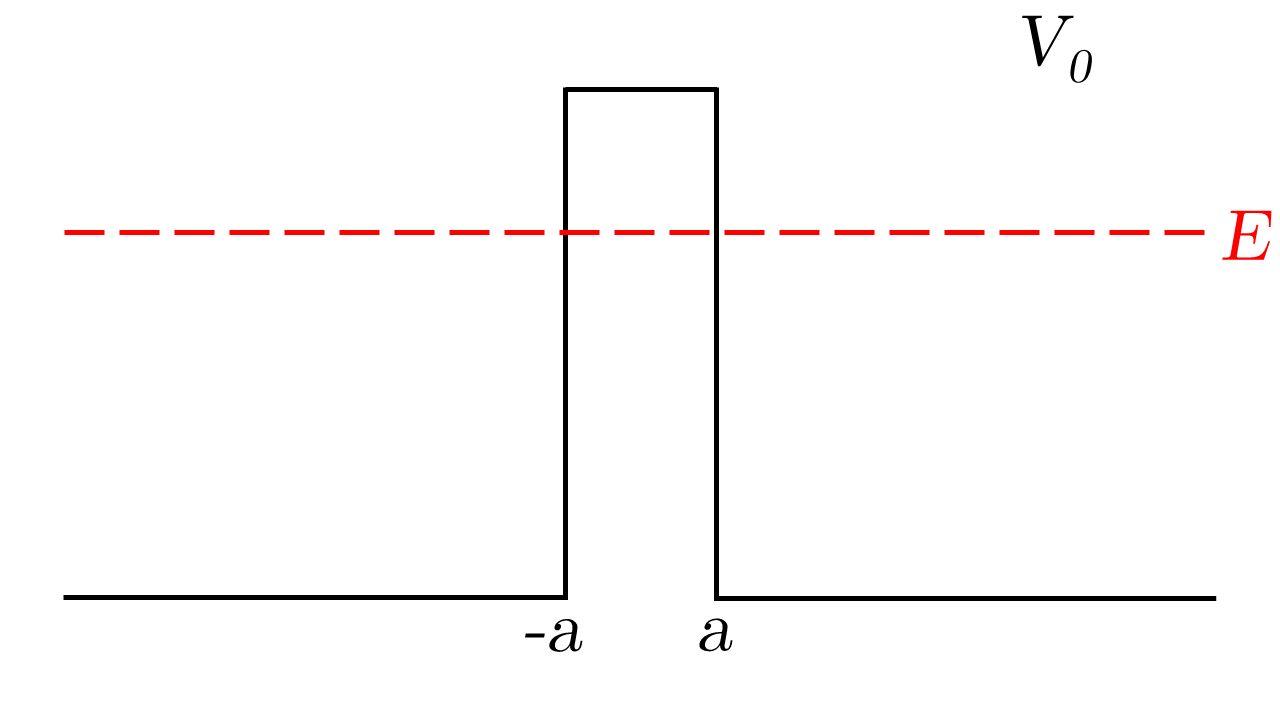}
			\caption{Tunneling through a potential}
			\label{fig2}
\end{figure}
Note that since we have assumed the Junction to be uniform in the transverse
directions, the system is effectively one dimension and so the current
is uniquely defined. From \eq{Jx} we obtain
\be
J =  \frac{z \zeta}{2 \zeta_0} \cdot \frac{ \hbar}{m \zeta_0}
{\rm Im}(C_1^* C_2).
\ee
The coefficients $C_1, C_2$ can be determined by imposing the boundary
conditions
%c5
that the wavefunction matches up with the Landau-Ginzburg order parameters
\eq{LGOP} of the
superconductors at $x = \pm a$:
\be\label{BCC}
\psi(-a) = \sqrt{n_1} e^{\ri \th_1}, \quad \psi(a) = \sqrt{n_2} e^{\ri \th_2}.
\ee
Taking into account the Cooper pair has charge $q = 2e$,
%c6
we obtain  the first  Josephson equation for the tunneling current
$\cJ := q J$:
\be \label{JJ1}
\cJ = \cJ_c \sin \vphi,\quad \vphi = \th_2 - \th_1.
\ee
Here the critical current  $\cJ_c$ is given by
\be
\cJ_c = \g \cJ_{c0}
\ee
with
\be
 \cJ_{c0} = \frac{ q \hbar \sqrt{n_1 n_2}}{m \zeta_0}\frac{1}{\sinh 2a/\zeta_0},
\quad
\g= \frac{z \zeta}{2 \zeta_0} \frac{\sinh 2 a/\zeta_0}{\sinh 2a/\zeta},
\ee
where $\cJ_{c0}$ is the critical current in the isotropic case and $\g$ is an
modification factor due to anisotropy.
If we consider the application of
a voltage $V$ across the Josephson Junction, then the phase becomes time
dependent and we have the second  Josephson equation,
  \be \label{JJ2}
\frac{\del \vphi}{\del t} = \frac{q}{\hbar} V.
\ee
Note that this result is unmodified by anisotropy.  
The derivation is the same as usual. In fact, due to gauge invariance,
the wavefunction in
the superconductor is modified to have a phase factor
\be
\psi \to \psi \exp\left( \frac{iq}{\hbar c} \int \bA \cdot d\bx\right)
\ee
when an electromagnetic field is introduced.
This results in a modification to the order parameter 
\be
\th_2 -\th_1 = (\th_2 -\th_1)_0 + \frac{q}{\hbar c} \int_1^2 \bA \cdot d\bx,
  \ee
  where the subscript ``0'' denotes the quantity before the electric field
  is introduced. 
This leads immediately to \eq{JJ2}.

Note that in the usual isotropic case,
the critical current $\cJ_{c0}$ is exponentially suppressed by
%c4
a factor of 
$e^{-2a/\zeta}$ since the Cooper pairs cannot tunnel
deep enough into the insulating region. However this may change in the
anisotropic situation. In fact, the  Josephson current can  be enhanced
($\g>1$) if the  penetration length is made bigger ($\zeta >\zeta_0$).
This happens  if
$z<2$ and $ \l_M < \zeta_0$ or $z>2$ and $ \l_M > \zeta_0$. Thus
an anisotropic material has the potential of enhancing the working
of Josephson Junction. In particular, if the anisotropic
insulating material can be
engineered such that the penetration length $\zeta$ get to become
of the same order as $a$,
then tunneling can become
much more effective and $\cJ_c$ may be
%c5 greatly
significantly enhanced.
As illustration, let us consider an isotropic sample with
$\zeta_0/2a =1/10$, in this
case the critical current is suppressed with a factor of
$\sim e^{-10} \sim 10^{-5}$. Introducing anisotropy, for example with
$z=0.8$ and $\zeta_0/\l_M = 10$, we can
have a huge amplification factor  $\g \sim  10^5$; or with
$z=5$ and $\zeta_0/\l_M = 1/10$, we have
$\g \sim  10^4$. We remark that
the dispersion relation \eq{dispersion} is acausal for $z<1$ due to the
existence of  superluminal modes \cite{Koroteev:2011zz} and the violation of
the null energy conditions of the holographic dual \cite{Hoyos:2010at}.
%c4
Therefore a junction with $z \geq 1$ may be more physical.
It is interesting to verify this prediction of enhanced tunneling
experimentally. 

The amplification of critical current due to time anisotropy
may be useful for improving the performance of Josephson Junction
in quantum computing.
Josephson Junction also found useful application in
digital circuits as switching elements due to it’s high
switching speed and low power dissipation \cite{Mukhanov2019}.
Some characteristic time constants in
Josephson Junction depends on the critical current, an
anisotropic Junction can have time constants that are
more useful for the switching purpose. Given the wide variety of
applications of  Josephson Junction \cite{Orlando:1991fas},
anisotropic Josephson Junction
appears to be an interesting topic that deserves further study.

\section*{Acknowledgments}
%c3
CSC would like to thank Harold Steinacker for discussions and
the physics department of the University of
Vienna for hospitality where part of the work was done.
The support of this work by NCTS and
the grant 113-2112-M-007-039-MY3 of the National 
Science and Technology Council of Taiwan is gratefully acknowledged.


\begin{thebibliography}{00}

\bibitem{lif}
  E. Lifshitz, \emph{On the theory of Second-Order Phase
  transitions I II}, Zh. Eksp.Teor. Fiz. 11
(1941) 255, 269.
  
  %\cite{Hornreich:1975zz}
\bibitem{Hornreich:1975zz}
R.~M.~Hornreich, M.~Luban and S.~Shtrikman,
\emph{Critical Behavior at the Onset of $\Vec{k}$-Space 
Instability on the $\lambda$ Line},
Phys. Rev. Lett. \textbf{35} (1975), 1678-1681.
%214 citations counted in INSPIRE as of 14 Jul 2025

\bibitem{Arkani-Hamed:2003pdi}
N.~Arkani-Hamed, H.-C.~Cheng, M.A.~Luty and S.~Mukohyama, \emph{{Ghost
  condensation and a consistent infrared modification of gravity}},
  \href{https://doi.org/10.1088/1126-6708/2004/05/074}{\emph{JHEP} {\bfseries
  05} (2004) 074} [\href{https://arxiv.org/abs/hep-th/0312099}{{\ttfamily
  hep-th/0312099}}].


\bibitem{PhysRevLett.61.2376}
D.S.~Rokhsar and S.A.~Kivelson, \emph{Superconductivity and the quantum
  hard-core dimer gas},
  \href{https://doi.org/10.1103/PhysRevLett.61.2376}{\emph{Phys. Rev. Lett.}
    {\bfseries 61} (1988) 2376}.
  
\bibitem{Henley_2004}
C.L.~Henley, \emph{From classical to quantum dynamics at
  rokhsar{\textendash}kivelson points},
  \href{https://doi.org/10.1088/0953-8984/16/11/045}{\emph{Journal of Physics:
  Condensed Matter} {\bfseries 16} (2004) S891}.
  
\bibitem{Ardonne_2004}
E.~Ardonne, P.~Fendley and E.~Fradkin, \emph{Topological order and conformal
  quantum critical points},
  \href{https://doi.org/10.1016/j.aop.2004.01.004}{\emph{Annals of Physics}
  {\bfseries 310} (2004) 493}.

\bibitem{Fradkin:2006mb}
E.~Fradkin and J.E.~Moore, \emph{{Entanglement entropy of 2D conformal quantum
  critical points: hearing the shape of a quantum drum}},
  \href{https://doi.org/10.1103/PhysRevLett.97.050404}{\emph{Phys. Rev. Lett.}
  {\bfseries 97} (2006) 050404}
  [\href{https://arxiv.org/abs/cond-mat/0605683}{{\ttfamily
  cond-mat/0605683}}].

\bibitem{Fradkin:2009dus}
E.~Fradkin, \emph{{Scaling of Entanglement Entropy at 2D quantum Lifshitz fixed
  points and topological fluids}},
  \href{https://doi.org/10.1088/1751-8113/42/50/504011}{\emph{J. Phys. A}
  {\bfseries 42} (2009) 504011}
  [\href{https://arxiv.org/abs/0906.1569}{{\ttfamily 0906.1569}}].

\bibitem{PhysRevB.80.184421}
J.-M.~St\'ephan, S.~Furukawa, G.~Misguich and V.~Pasquier, \emph{Shannon and
  entanglement entropies of one- and two-dimensional critical wave functions},
  \href{https://doi.org/10.1103/PhysRevB.80.184421}{\emph{Phys. Rev. B}
  {\bfseries 80} (2009) 184421}.

\bibitem{Hsu:2008af}
B.~Hsu, M.~Mulligan, E.~Fradkin and E.-A.~Kim, \emph{{Universal entanglement
  entropy in 2D conformal quantum critical points}},
  \href{https://doi.org/10.1103/PhysRevB.79.115421}{\emph{Phys. Rev. B}
  {\bfseries 79} (2009) 115421}
  [\href{https://arxiv.org/abs/0812.0203}{{\ttfamily 0812.0203}}].

\bibitem{Hsu:2010ag}
B.~Hsu and E.~Fradkin, \emph{{Universal Behavior of Entanglement in 2D Quantum
  Critical Dimer Models}},
  \href{https://doi.org/10.1088/1742-5468/2010/09/P09004}{\emph{J. Stat. Mech.}
  {\bfseries 1009} (2010) P09004}
  [\href{https://arxiv.org/abs/1006.1361}{{\ttfamily 1006.1361}}].

\bibitem{Oshikawa:2010kv}
M.~Oshikawa, \emph{{Boundary Conformal Field Theory and Entanglement Entropy in
  Two-Dimensional Quantum Lifshitz Critical Point}},
[\href{https://arxiv.org/abs/1007.3739}{{\ttfamily 1007.3739
      }}].

\bibitem{PhysRevLett.107.020402}
M.P.~Zaletel, J.H.~Bardarson and J.E.~Moore, \emph{Logarithmic terms in
  entanglement entropies of 2d quantum critical points and shannon entropies of
  spin chains},
  \href{https://doi.org/10.1103/PhysRevLett.107.020402}{\emph{Phys. Rev. Lett.}
  {\bfseries 107} (2011) 020402}.
       [\href{https://arxiv.org/abs/1103.5452}{{\ttfamily 1103.5452
             }}].
       
\bibitem{Zhou:2016ykv}
T.~Zhou, X.~Chen, T.~Faulkner and E.~Fradkin, \emph{{Entanglement entropy and
  mutual information of circular entangling surfaces in the 2 + 1-dimensional
  quantum Lifshitz model}},
  \href{https://doi.org/10.1088/1742-5468/2016/09/093101}{\emph{J. Stat. Mech.}
  {\bfseries 1609} (2016) 093101}
  [\href{https://arxiv.org/abs/1607.01771}{{\ttfamily 1607.01771}}].

\bibitem{MohammadiMozaffar:2017nri}
M.R.~Mohammadi~Mozaffar and A.~Mollabashi, \emph{{Entanglement in Lifshitz-type
  Quantum Field Theories}},
  \href{https://doi.org/10.1007/JHEP07(2017)120}{\emph{JHEP} {\bfseries 07}
  (2017) 120} [\href{https://arxiv.org/abs/1705.00483}{{\ttfamily
  1705.00483}}].


\bibitem{Berthiere:2019lks}
C.~Berthiere and W.~Witczak-Krempa, \emph{{Relating bulk to boundary
  entanglement}},
  \href{https://doi.org/10.1103/PhysRevB.100.235112}{\emph{Phys. Rev. B}
  {\bfseries 100} (2019) 235112}
       [\href{https://arxiv.org/abs/1907.11249}{{\ttfamily 1907.11249}}].
       
\bibitem{Angel-Ramelli:2019nji}
J.~Angel-Ramelli, V.G.M.~Puletti and L.~Thorlacius, \emph{{Entanglement Entropy
  in Generalised Quantum Lifshitz Models}},
  \href{https://doi.org/10.1007/JHEP08(2019)072}{\emph{JHEP} {\bfseries 08}
  (2019) 072} [\href{https://arxiv.org/abs/1906.08252}{{\ttfamily
  1906.08252}}].

 
\bibitem{Angel-Ramelli:2020wfo}
J.~Angel-Ramelli, C.~Berthiere, V.G.M.~Puletti and L.~Thorlacius,
  \emph{{Logarithmic Negativity in Quantum Lifshitz Theories}},
  \href{https://doi.org/10.1007/JHEP09(2020)011}
       {\emph{JHEP} {\bfseries 09}
  (2020) 011} [\href{https://arxiv.org/abs/2002.05713}{{\ttfamily
        2002.05713}}].
 
\bibitem{Berthiere:2023bwn}
C.~Berthiere, B.~Chen and H.~Chen, \emph{{Reflected entropy and Markov gap in
  Lifshitz theories}},
  \href{https://doi.org/10.1007/JHEP09(2023)160}{\emph{JHEP} {\bfseries 09}
  (2023) 160} [\href{https://arxiv.org/abs/2307.12247}{{\ttfamily
        2307.12247}}].

  
\bibitem{Keranen:2016ija}
V.~Keranen, W.~Sybesma, P.~Szepietowski and L.~Thorlacius,
\emph{{Correlation functions in theories with Lifshitz scaling}},
 \href{https://link.springer.com/article/10.1007/JHEP05(2017)033}
{\emph{JHEP }{\bfseries 05} (2017), 033}
[\href{https://arxiv.org/abs/1611.09371}{{\ttfamily
      1611.09371 }}].
  
%\cite{Basak:2023otu}
\bibitem{Basak:2023otu}
J.~K.~Basak, A.~Chakraborty, C.~S.~Chu, D.~Giataganas and H.~Parihar,
\emph{Massless Lifshitz field theory for arbitrary z},
 \href{https://doi.org/10.1007/JHEP05(2024)284}
{\emph{JHEP} \textbf{05} (2024), 284},
  [\href{https://arxiv.org/abs/2312.16284} {{\ttfamily 2312.16284}}].

 \bibitem{Chu:2024nwf}
C.~S.~Chu, I.~G.~Gonzalez and H.~Parihar,
\emph{Holography for boundary Lifshitz field theory},
\href{https://doi.org/10.1007/JHEP11(2024)158}
{\emph{JHEP} \textbf{11} (2024), 158},
  [\href{https://arxiv.org/abs/2409.06667}{{\ttfamily 2409.06667}}].

%\cite{deAlfaro:1976vlx}
\bibitem{deAlfaro:1976vlx}
V.~de Alfaro, S.~Fubini and G.~Furlan,
``Conformal Invariance in Quantum Mechanics,''
Nuovo Cim. A \textbf{34} (1976), 569.
%655 citations counted in INSPIRE as of 20 Jul 2025


%\cite{Laskin:2010ry}
    \bibitem{Laskin:2010ry}
    N.~Laskin,
    \emph{Principles of Fractional Quantum Mechanics},
    World Scientific, 2011.
    [\href{https://arxiv.org/abs/1009.5533}{{\ttfamily math-ph/1009.5533}}].
    %13 citations counted in INSPIRE as of 10 Nov 2024
    
%\cite{Tayurskii2012}
    \bibitem{Tayurskii2012}
    D.~A.~Tayurskii and Yu.~V.~Lysogorskiy, 
    \emph{Superfluid Hydrodynamics in Fractal Dimension Space}, 
           \href{https://dx.doi.org/10.1088/1742-6596/394/1/012004}{\emph{Journal
               of Physics: Conference Series}, 394 (2012) 012004}. 

%\cite{Wei2016}
    \bibitem{Wei2016}
    Y.~Wei, 
    \emph{Comment on ``Fractional quantum mechanics'' and
      ``Fractional Schrödinger equation''},  
    \href{https://link.aps.org/doi/10.1103/PhysRevE.93.066103}
{\emph{Phys. Rev. E}, 93 (2016) 066103}.




  
%\cite{Laskin2016}
    \bibitem{Laskin2016}
    N.~Laskin, 
    \textit{Reply to ``Comment on `Fractional quantum mechanics' and
      `Fractional Schrödinger equation'''}, 
    Phys. Rev. E, 93 (2016) 066104.  
    %  [\textcolor{blue}{\url{https://link.aps.org/doi/10.1103/PhysRevE.93.066104}}].


   \bibitem{Koroteev:2011zz}
P.~Koroteev, \emph{{Causality and Lifshitz holography}},
   Nucl. Phys. B
  Proc. Suppl. 216 (2011) 245.

\bibitem{Hoyos:2010at}
C.~Hoyos and P.~Koroteev, \emph{{On the Null Energy Condition and Causality in
  Lifshitz Holography}},
  \href{https://doi.org/10.1103/PhysRevD.82.109905}{\emph{Phys. Rev. D}
  {\bfseries 82} (2010) 084002}
  [\href{https://arxiv.org/abs/1007.1428}{{\ttfamily 1007.1428}}].

%\cite{Mukhanov2019}
    \bibitem{Mukhanov2019}
    O.~Mukhanov, N.~Yoshikawa, I.~P.~Nevirkovets, and M.~Hidaka,
    \textit{Josephson Junctions for Digital Applications}, in 
    F.~Tafuri, editor, \textit{Fundamentals and Frontiers of the
      Josephson Effect}, 
    \href{https://doi.org/10.1007/978-3-030-20726-7_16}{\emph{
      Springer International Publishing, Cham} (2019), pp.~611--701.}
   


%\cite{Orlando:1991fas}
    \bibitem{Orlando:1991fas}
    T.~P.~Orlando and K.~A.~Delin,
    \textit{Foundations of Applied Superconductivity},
    Addison-Wesley (1991).

\end{thebibliography}
\end{document}